\documentstyle[prl,aps]{revtex}
\begin{document}
\draft
\title{New critical behaviour of the three-dimensional Ising model with
nearest-neighbor, next-nearest-neighbor and plaquette interactions}
\author{Emilio N.M. Cirillo, G. Gonnella}
\address{Dipartimento di Fisica dell'Universit\`{a} di Bari and
Istituto Nazionale di Fisica Nucleare, Sezione di Bari
via Amendola 173, 70126 Bari, Italy}
\author{A. Pelizzola}
\address{Dipartimento di Fisica del Politecnico di Torino and
Istituto Nazionale per la Fisica della Materia,
c. Duca degli Abruzzi 24, 10129 Torino, Italy}
\date{\today}
\maketitle
\begin{abstract}
The critical and multicritical behavior of the simple cubic Ising
model with nearest-neighbor, next-nearest-neighbor and plaquette
interactions is studied using the cube and star-cube approximations of
the cluster variation method and the recently proposed cluster
variation--Pad\'e approximant method. Particular attention is paid to
the line of critical end points of the ferromagnetic-paramagnetic
phase transition: its (multi)critical exponents are calculated, and
their values suggest that the transition belongs to a novel
universality class. A rough estimate of the crossover exponent is also
given. 
\end{abstract}

\pacs{PACS numbers: 05.50.+q; 64.60.Fr; 68.35.Rh}

\narrowtext

The Ising model was introduced in the 20's  \cite{Is} for describing 
magnetic transitions and  is still the subject of a very intense
research activity. In particular, the Ising model with
nearest-neighbor (n.n.), next-nearest-neighbor (n.n.n.) and plaquette
interactions on the simple cubic lattice has been considered as a
simple model for the statistical mechanics of random surfaces
\cite{NPW,m1,Kar}, microemulsions \cite{DG}, and also as a discretized
string action (the so-called gonihedric model)
\cite{Savv,WS,des,pw,pla}. The critical behavior of the square lattice
version of this model has been studied for many years and is now
well-established \cite{2d}, but very little has been done in the
three-dimensional case. Previous mean-field calculations \cite{m1,m2}
have shown that the model exhibits a very rich phase diagram, with
lamellar and ordered bicontinuous \cite{scri} phases, 
disordered structured and non-structured regions \cite{wid,cgm},
and coexisting ferromagnetic and paramagnetic phases with a  first-order
wetting transition \cite{GS}.

The purpose of the present paper is to investigate the critical
properties of the model beyond the mean field level \cite{sim}, focusing mainly
on the multicritical and crossover properties of the line of critical
end points of the ferromagnetic-paramagnetic phase transition. This
will be done by means of the cluster variation method (CVM)
\cite{kik1,an} in its cube and star-cube \cite{ap-physa}
approximations, and the recently proposed cluster variation--Pad\'e
approximant method (CVPAM) \cite{ap-rc,ap-jmmm,ap-pre}. Before turning
to the description of our results, we now give a short account of both
the model and the method. 

The model is defined by the reduced 
hamiltonian ${\cal H} = -\beta H$
\begin{equation}
{\cal H} = J_1 \sum_{<ij>} s_i s_j + J_2\sum_{<<ij>>} s_i s_j
          +J_3 \sum _{[i,j,k,l]} s_i s_j s_k s_l,
\label{vattelapesca}
\end{equation}
where $\sigma_i = \pm 1$ is the Ising variable associated to the site
$i$ of our simple cubic lattice, and sums are respectively over the
n.n.\ pairs, n.n.n.\ pairs and plaquettes of the lattice. 

In terms of the Peierls surfaces separating domains of spins with different 
sign, not only the area but also the bending and the intersections
of the Peierls interfaces are weighted by the couplings $J_1,J_2,J_3$ 
\cite{Kar}. 
If $\beta_A, \beta_C, \beta_I$ are respectively the
energy cost for a plaquette, a bending between two adjacent
plaquettes and an intersection of four plaquettes sharing a common
dual link of the Peierls interfaces, the relation with
the couplings $J_1, J_2, J_3$ is $\beta_A = 2 J_1 + 8 J_2$, $\beta_C
= 2 (J_3 - J_2)$, $\beta_I = - 4 (J_2 + J_3)$ \cite{m1}.
The model (\ref{vattelapesca}) can be considered as 
a discrete realization of a random surface model
with an extrinsic curvature energy term \cite{Hel,Pol}. 
Recently, the special case in which
only bendings and intersections are taken into account 
and the area is  not weighted at all (that is $J_1/\beta= 2 \kappa$,
$J_2/\beta= - \kappa/2$, $J_3/\beta=(1-\kappa)/2$, where $\kappa = 0$ is
the case of phantom surfaces \cite{pha}, while $\kappa \to \infty$
represents the 
limit of complete self-avoidance) has been put in connection
with a discretized string model (the so-called gonihedric model)
\cite{Savv} and studied by Savvidy and Wegner 
\cite{WS}. This choice of the couplings corresponds  to a
zero temperature 
high degeneracy point where all possible sequences of
``$+$'' and ``$-$'' planes have the same energy.
A phase transition has been found in this restricted
parameter space with exponents different from the usual
$3d$ Ising exponents \cite{des,pw,pla}.
It has to be observed that this special case
corresponds to the disorder line \cite {disli} $J_2=- J_1/4$ 
as calculated in the mean-field approximation \cite{m2} and that in the two
dimensional case there is no transition neither on the disorder line \cite{bax}
nor on the line $J_2=-J_1/4$ \cite{Bind}.

We will concentrate on the study of the phase diagram 
around the region where the lamellar, the ferromagnetic and the 
paramagnetic phases coexist, which is close to the line $J_2=-J_1/4$,
and we will explain the origin of the transition found in \cite{des,pw}.
There is a particular physical interest in this region due to
the extremely low values of the surface tension between coexisting
phases \cite{m2}, 
which is an important property for applications in real surfactant systems
\cite{DG}.

Let us now briefly discuss the methods we are going to use. The
cluster variation method is a powerful generalized mean field theory
introduced by Kikuchi \cite{kik1} and then reformulated in a very
elegant way \cite{an} as a truncated cluster (cumulant) expansion of
the variational principle of statistical mechanics. In the cube
approximation of the CVM one has to minimize 
the free energy density functional 
\begin{eqnarray}
f[\rho_8] &=& {\rm Tr} (\rho_8 H_8) + \frac{1}{\beta} \Bigg[
{\rm Tr} {\cal L} (\rho_8) 
- \frac{1}{2} \sum_{\rm plaqs} 
{\rm Tr} {\cal L} (\rho_{4,{\rm plaq}}) \nonumber \\
&& + \frac{1}{4} \sum_{\rm edges}
{\rm Tr} {\cal L} (\rho_{2,{\rm edge}}) - \frac{1}{8} \sum_{\rm sites}
{\rm Tr} {\cal L} (\rho_{1,{\rm site}}) \Bigg],
\label{fcube}
\end{eqnarray}
where $H_8$ is the contribution of a single cube to the hamiltonian
(when splitting the total hamiltonian $H$ into single cube
contributions one has to keep in mind that nearest neighbour
interactions are shared by four cubes 
and then will get a coefficient 1/4 in $H_8$, and similarly
next-nearest neighbour and plaquette interactions will get a
coefficient 1/2), ${\cal L}(x) = x \ln x$, $\rho_\alpha$ with $\alpha
= 8$ (4, 2, 1) denotes the cube (respectively plaquette, edge, site)
density matrix, and the sums in the entropy part are over all
plaquettes (edges, sites) of a single cube (notice that we have not
assumed any {\em a priori} symmetry property for our density matrices,
and that the plaquette, edge and site matrices can be thought of as
partial traces of the cube matrix). In the following we shall also
use, for the ferromagnetic phase only, the star-cube approximation,
introduced in \cite{ap-physa}, where it is described in great detail
for the n.n.\ case (inclusion of n.n.n.\ and plaquette interactions is
indeed straightforward). For both approximations, the (numerical)
minimization task is greatly simplified by the so-called natural
iteration method \cite{ap-physa,kik2}.

Being an approximate variational theory, the CVM yields necessarily
classical values of the critical exponents. In order to overcome this
major drawback, one of us has proposed the cluster variation--Pad\'e
approximant method \cite{ap-rc,ap-jmmm,ap-pre}. The basic idea of the
CVPAM is that, since the CVM with 7-8 point or larger clusters is very
accurate at high and low enough temperatures, one can extrapolate the
results at such temperatures using Dlog Pad\'e approximants
\cite{guttmann} (see
\cite{ap-pre} for an application of the more sophisticated Adler's
methods) in order to extract accurate information about the critical
behavior, i.e.\ improved critical temperatures and non-classical,
precise critical exponents. In test applications, indeed, the CVPAM
has produced results of quality almost comparable to state of the art
Monte Carlo simulations, but with a much smaller numerical effort. 

We can describe  our results. The ground states of
the model have been thoroughly investigated in \cite{m1}; here we shall
consider the restricted  
parameter space  $J_1, J_2$ and  $\kappa >0 $ with $J_3/J_1=
(1-\kappa)/(4\kappa)$. In this region, the line 
$J_2/J_1 = -1/4$ is always the boundary between
 ferromagnetic and lamellar ground states.

In Fig.\ \ref{diag} the phase diagram of the model
(\ref{vattelapesca}) as given by the CVM cube approximation is
depicted in the plane $J_3=0$. The line separating the paramagnetic and the 
ferromagnetic phase (dashed line) is a second order line, with the usual 
critical Ising transition at $J_1=0.218$ (shifted to $J_1 = 0.222$ by
application of the CVPAM \cite{ap-rc}), to be compared with the best estimate
$J_1=0.22165$\cite{Lan}. The lamellar 
phase - here consisting of alternate planes of different sign - 
is separated from the paramagnetic and the ferromagnetic phase by a 
coexistence line (solid line) \cite{caz}, which is asymptotically close to the 
line $J_2=-J_1/4$ at low temperature. 
The second order line ends onto the first order 
one with a critical end point at 
$J_1^{\rm end}=0.865\pm 0.005$ and $J_2^{\rm end}=-0.2176\pm 0.0006$.
In the case of the gonihedric model \cite{Savv} 
with $\kappa = 1$, our value for the inverse critical temperature
is $\beta_c = 0.427$ to be compared with the value $\beta_c=0.44$
found by Monte Carlo simulations \cite{des} and the upper bound
$\beta_c=1.49$ calculated in \cite{pw}. Notice that, since the
lamellar-ferromagnetic coexistence line is slightly bent toward the
lamellar phase (and this feature persists at $J_3 \ne 0$), the
critical point of the gonihedric model is extremely close to our
critical end point, and this has important consequences on the meaning
of the exponents that one can define, as we shall see. 

The topology of the phase diagram at varying  $\kappa$ 
remains 
the same as at $J_3=0$, but in the range $0 < \kappa < \kappa^*$, with
$\kappa^* \simeq 0.8$,  
there is a tricritical point (see Fig.\ \ref{fig2}) on the
ferromagnetic-paramagnetic 
transition line at negative values
of $J_2$ \cite{next}. 
The transition line becomes of first order before reaching the 
lamellar phase. 

The critical behaviour
suggested by the phase diagram of Fig. \ref{diag} can be now discussed.
The first important point is that the 
 critical end point we have found (actually a
line of critical end points, since $\kappa$ can vary),
 which does not exist in
two dimensions \cite{2d},
must be described by critical exponents
which differ from the usual three-dimensional Ising ones, which apply
to the ferromagnetic-paramagnetic critical surface. As a consequence,
in the vicinity of the line of critical end points, which bounds the
critical surface, it is natural to expect some crossover phenomenon. 

In Ref.\ \cite{pla} we have already calculated with the CVPAM the
order parameter 
critical exponent $\beta$ of the gonihedric model (using CVM results
up to a temperature which was less than half the transition
temperature), finding $\beta = 
0.062 \pm 0.003$ (together with the improved estimate for the inverse
critical temperature $\beta_c = 0.434$), which agrees well with the
Monte Carlo estimates $\beta/\nu = 0.04(1)$ and $\nu = 1.2(1)$
\cite{des}. In view of the above considerations this must be 
regarded as an effective exponent (in our picture the critical
transition of the gonihedric model lies on the universal
ferromagnetic-paramagnetic critical surface, and hence has to be
described by the usual 3d Ising exponents), induced by a crossover
phenomenon. Nevertheless, the critical
transition of the gonihedric model is extremely close to our critical
end point and this means that the corresponding exponents are very
good approximations to the critical end point ones. Thus from now on
we shall use the estimate $\beta_{\rm CEP} = 0.062 \pm 0.003$. In order
to calculate also  $\gamma_{\rm CEP}$, 
we have determined the high
temperature susceptibility for $J_3 = 0$ and $J_2 = - J_1/4$ in the
cube approximation of the CVM and, according to the CVPAM
prescriptions, we have determined Dlog Pad\'e approximants (biased
with our improved $\beta_c$) to it, and from these we have deduced
$\gamma_{\rm CEP} = 1.41 \pm 0.02$. 

The analysis of the crossover phenomenon is a considerably more
difficult task, and we have tried to give an estimate of the crossover
exponent $\phi$ \cite{cross} proceeding along the lines described in
\cite{ap-jmmm}. From now on we set $J_2/J_1 = R$.
Assuming that near the critical end point, but
still in the ferromagnetic phase, the order
parameter has a multicritical scaling law given by 
\begin{equation}
m \simeq t^{\beta_{\rm CEP}} f(z), \qquad z = \frac{R - R_{\rm
CEP}}{t^\phi},
\label{m-scal}
\end{equation}
where $t$ is the deviation from the critical temperature and $R_{\rm
CEP}$ can be well approximated by $-1/4$ (see also below), one can
derive the scaling laws 
\begin{equation}
T_c(R) - T_c(R_{\rm CEP}) \propto (R - R_{\rm CEP})^{1/\phi}
\label{tc-scal}
\end{equation}
for the ferromagnetic critical temperature and
\begin{equation}
B(R) \propto (R - R_{\rm CEP})^{-\omega}, \qquad \omega =
\frac{\beta_{\rm Ising} - \beta_{\rm CEP}}{\phi},
\label{amp-scal}
\end{equation}
where $\beta_{\rm Ising} \simeq 0.327$ \cite{blotal} is the usual
three-dimensional Ising exponent (a high-order CVPAM analysis on the
simple cubic lattice predicted $\beta_{\rm Ising} = 0.325(4)$
\cite{ap-pre}), for the critical amplitude $B(R)$, 
which is defined by
\begin{equation}
m \simeq B(R)(T_c(R) - T)^{\beta_{\rm Ising}}, \qquad R > R_{\rm CEP}.
\label{amp-def}
\end{equation}

Using the CVPAM we have extrapolated the low-temperature ferromagnetic order
parameter (the smallest value used being 0.89) given by the CVM
star-cube approximation, to 
calculate the critical temperatures and amplitudes in the range $-0.24
\le R \le -0.14$ for $\kappa = 1, 2$ and 10. $T_c(R)$ was determined by
requiring that $T_c(R)$--biased Dlog
Pad\'e approximants gave 
$\beta_{\rm Ising} = 0.327$ and then the critical amplitude could be
obtained by making approximants to $(T_c(R) - T)^{-\beta_{\rm Ising}} m$.
As a check, for the simple n.n.\ Ising model (that is $\kappa = 0$ and
$R = 0$) we have obtained $B(0) \simeq 1.626$, to be compared with the
value 1.6919045 reported by Bl\"ote and Talapov. The results for
$\kappa = 1$ are reported in Tab.\ \ref{table}. Several fits were
then made. The fits on the basis of Eq.\ \ref{tc-scal} gave $\phi =
1.13, 1.09$ and 1.03 for $\kappa = 1, 2$ and 10 respectively, while
the fits to Eq.\ \ref{amp-scal} gave $\phi = 1.16$ and 1.20 for
$\kappa = 1$ and 2 and were inconclusive for $\kappa = 10$. A fit with
$R_{\rm CEP}$ free was also made to Eq.\ \ref{amp-scal} for $\kappa =
1$, and the result was $R_{\rm CEP} = -0.249774$, confirming that
$R_{\rm CEP} = -1/4$ is a very good approximation. 

A reasonable final estimate for the crossover exponent might then be
$\phi = 1.1(1)$, but it must be taken with some care for several
reasons. Apart from the various approximations involved in the
calculations of $\phi$, it is a matter of fact that a similar
calculation in the case of the semi-infinite Ising model
\cite{ap-jmmm} gave a result differing by 10 to 30\% from extensive
computer simulations, and furthermore the true multicritical scaling
law for the order parameter might be more complicated than our Eq.\
\ref{m-scal} (e.g.\ the scaling axes might not be parallel to the $T$
and $R$ axes), although the relatively good quality of the fits seems
to indicate that Eq.\ \ref{m-scal} is fairly good. 

Summarizing, we have studied the critical and multicritical behavior
of the simple cubic Ising model with n.n., n.n.n.\ and plaquette
interactions, calculating the exponents of a line of critical end
points which does not exist in two dimensions and might be relevant
for some surfactant system. 
We have also tried to give an estimate of the crossover
exponent, but this can certainly be refined using e.g.\ Monte Carlo
simulations or series expansions combined with the partial
differential approximants method \cite{pda}.

\acknowledgements

We are indebted to A. Maritan, D.A. Johnston and M. Caselle
for valuable discussions.

\mediumtext

\begin{table}
\caption{Inverse critical temperatures $J_c(R) = 1/T_c(R)$ and order
parameter amplitudes for $\kappa = 1$. 
\label{table}}
\begin{tabular}{cccccccccccc}
$R$ & -0.14 & -0.15 & -0.16 & -0.17 & -0.18 & -0.19 & -0.20 & -0.21 &
-0.22 & -0.23 & -0.24 \\
$J_{1,c}(R)$ & 0.360 & 0.378 & 0.398 & 0.421 & 0.447 & 0.476 &
0.511 & 0.552 & 0.602 & 0.663 & 0.743 \\
$B(R)$ & 1.84 & 1.87 &1.90& 1.95 & 2.01 & 2.06 & 2.17 & 2.28 & 2.47 &
2.68 & 3.15 \\
\end{tabular}
\end{table}

\narrowtext

\begin{figure}
\caption{The phase diagram of the model (\protect\ref{vattelapesca})
for $J_3=0$. Solid and the dashed lines represent first and
second order transitions respectively.}
\label{diag}
\end{figure}

\begin{figure}
\caption{The phase diagram of the model (\protect\ref{vattelapesca})
for $\kappa=1/3$. Solid and the dashed lines represent first and
second order transitions respectively.}
\label{fig2}
\end{figure}


\begin{references}
\bibitem{sim} Actually, Monte Carlo simulations have been 
performed \cite{m1,Kar}
to investigate the critical behaviour of model (\ref{vattelapesca}),
but only in the region close to the usual Ising transition point,
where tricritical points with classical exponents have been found
on the ferromagnetic-paramagnetic transition line.
\bibitem {pha} Surfaces without excluded volumes are sometimes called
phantom surfaces.
\bibitem{caz} Our calculations show that the ferromagnetic-lamellar 
transition line, even if quite close, does not exactly coincide
with the line $J_2=-J_1/4$ as it was assumed in \cite{des}.
\bibitem{next} More extensive studies of  the phase diagram
and of the tricritical end point
will appear elsewhere.
\end{references}
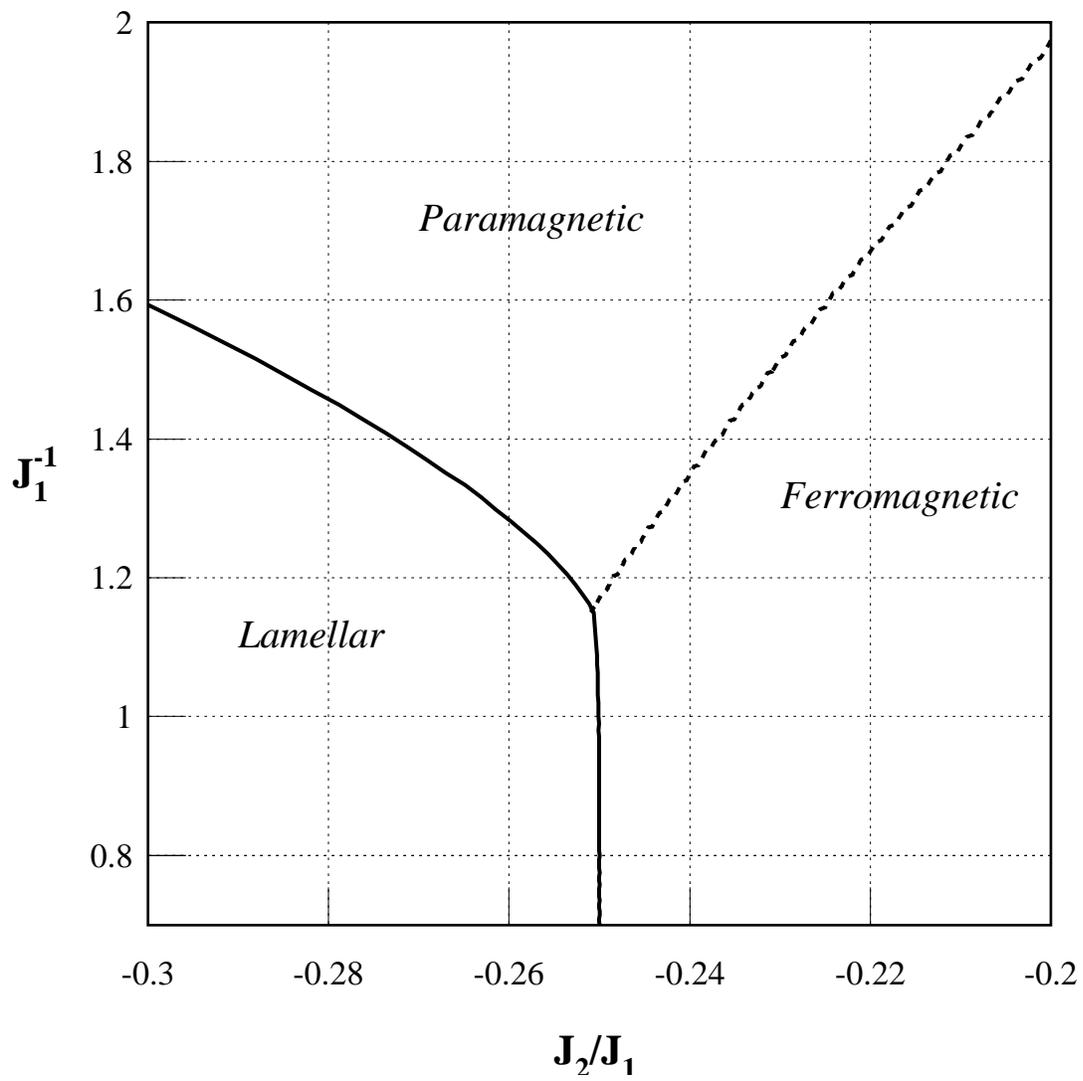
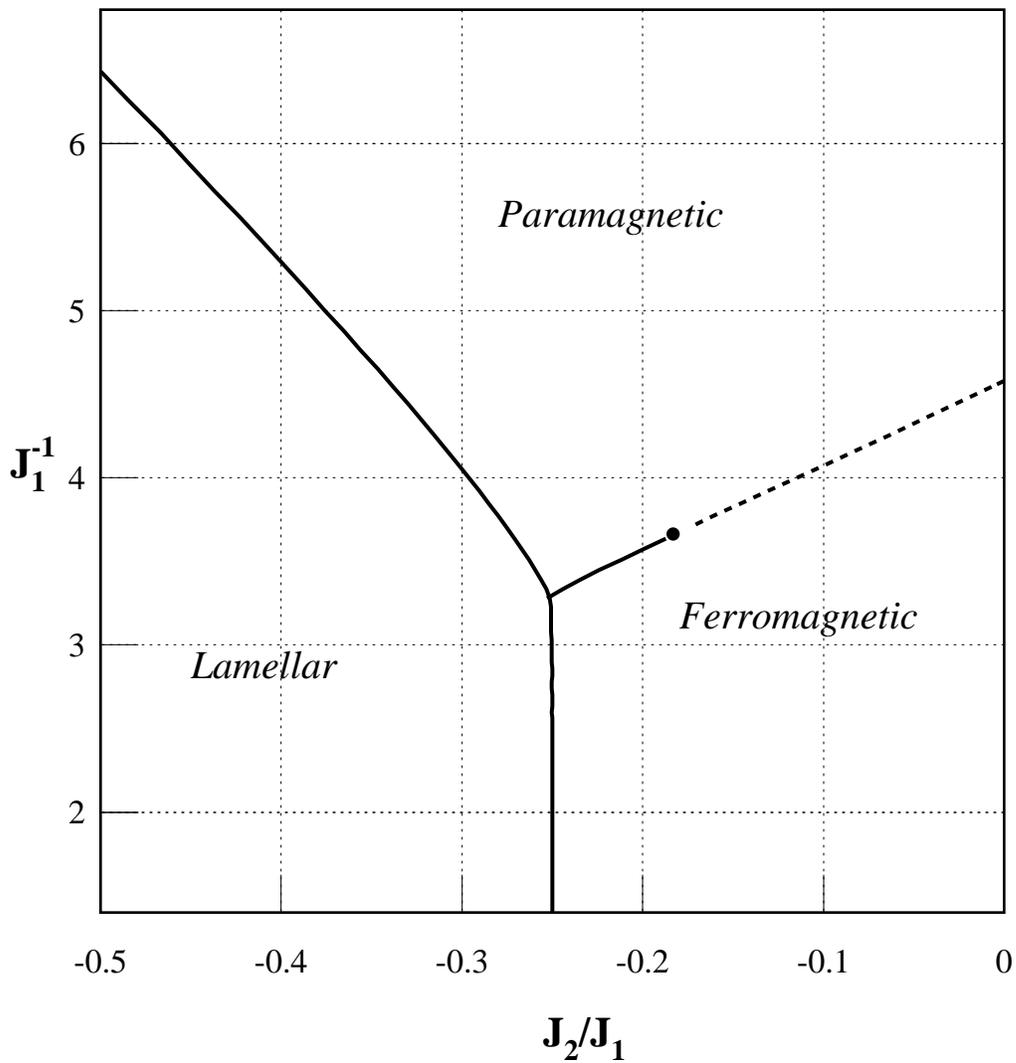
\end{document}